\documentclass[aps,pra,showpacs,groupedaddress,twocolumn]{revtex4}
\usepackage{graphicx}

\def\bh{\mathbf{h}}
\def\bH{\mathbf{H}}
\def\bhs{\bh^{*}}

\def\bHs{\bH^{*}}
\def\hi{h_i}
\def\his{\hi^*}

\def\hsiunc{\Delta\his}
\def\hiss{\hi\!\!\stackrel{\raisebox{-8pt}[-1.0ex][-0.0ex]{*}}{\raisebox{-
1.0pt}{*}}}\def\hissunc{\Delta\hiss}

\def\hsimin{^{<}\!\his}

\def\hsimax{^{>}\!\!\his}
\def\hissmin{^<\!\hi\!\!\!\stackrel{\raisebox{-8pt}[-1.0ex][-
0.0ex]{*}}{\raisebox{-1.0pt}{*}}}
\def\hissmax{^>\!\hi\!\!\!\stackrel{\raisebox{-8pt}[-1.0ex][-
0.0ex]{*}}{\raisebox{-1.0pt}{*}}}

\def\bc{\mathbf{c}}

\def\phio{\Phi^{(lab)}}

\def\phiok{\phio_k}
\def\phioks{\Phi^*_k}
\def\phioki{\phio_{k,m}}
\def\phick{\Phi_{k}[\mathbf{h}]}
\def\phicki{\Phi_{k,m}[\mathbf{h}]}
\def\phierr{\varepsilon^{(lab)}}

\def\phierrk{\phierr_k}
\def\phierrki{\phierr_{k,m}}

\def\hdim{N_h}

\def\hi{h_i}
\def\hj{h_j}

\def\hone{h_1}
\def\hN{h_{\hdim}}

\def\hiba\def\hbar{\bar{\bh}}

\def\bh{\mathbf{h}}

\def\bH{\mathbf{H}}
\def\bhs{\bh^{*}}

\def\bHs{\bH^{*}}

\def\hi{h_i}

\def\hssi{\hi\!\!\stackrel{\raisebox{-8pt}[-1.0ex][-0.0ex]{*}}{\raisebox{-1.0pt}{*}}}

\def\hssunci{\Delta\hssi}

\def\hssmini{^<\!\hi\!\!\!\stackrel{\raisebox{-8pt}[-1.0ex][-0.0ex]{*}}{\raisebox{-1.0pt}{*}}}

\def\hssmaxi{^>\!\hi\!\!\!\stackrel{\raisebox{-8pt}[-1.0ex][-0.0ex]{*}}{\raisebox{-1.0pt}{*}}}

\def\fh0{f_0}
\def\fhi{f_i(\hi;E_k)}

\def\fhij{f_{ij}(\hi,\hj;E_k)}
\def\fhlast{f_{1\cdots \hdim}(\hone,\ldots,\hN;E_k)}

\def\bc{\mathbf{c}}

\def\mapord{L}

\def\ie{\textit{i.e.}}

\def\etc{\textit{etc.}}
\def\etcend{\textit{etc}}

\def\reffig#1{Figure \ref{Figure::#1}}

\def\refeqn#1{Eq.\ (\ref{Equation::#1})}
\def\refeqs#1#2{Eqs.\ (\ref{Equation::#1}) and (\ref{Equation::#2})}

\def\sci#1#2{#1$\times$10$^{#2}$}

\def\ket#1{$| #1 \rangle$}

\begin{document}

\title{Efficient extraction of quantum Hamiltonians
from optimal laboratory data}

\author{JM Geremia}\email{jgeremia@Caltech.EDU}
\affiliation{Physics and Control \& Dynamical Systems, California
Institute of Technology, Pasadena, CA 91125}
\author{Herschel A. Rabitz}
\affiliation{Department of Chemistry, Princeton University,
    Princeton, NJ 08540}

\begin{abstract}
Optimal Identification (OI) is a recently developed procedure for
extracting optimal information about quantum Hamiltonians from
experimental data using shaped control fields to drive the system
in such a manner that dynamical measurements provide maximal
information about its Hamiltonian.  However, while optimal, OI is
computationally expensive as initially presented.  Here, we
describe the unification of OI with highly efficient global,
nonlinear map-facilitated data inversion procedures.  This
combination is expected to make OI techniques more suitable for
laboratory implementation. A simulation of map-facilitated OI is
performed demonstrating that the input-output maps can greatly
accelerate the inversion process.
\end{abstract}
\date{\today}
\maketitle

\section{Introduction}

A general goal in atomic and molecular physics is to
quantitatively predict quantum dynamics from knowledge of the
system Hamiltonian. However, sufficiently accurate information
about these Hamiltonians is still lacking for many applications.
Although the capabilities of \textit{ab initio} methods are
improving, they remain unable to provide the quantitative accuracy
needed to predict many quantum dynamical phenomena and data
inversion remains the most reliable source of precision
information about quantum Hamiltonians.  But, traditional data
inversion techniques are hindered by the fact that (a)
spectroscopic and collision data only provide information about
limited portions of the the desired interactions and (b) the
relationships between quantum Hamiltonians and their corresponding
observables are generally nonlinear \cite{GeremiaForwardMap}.

The recently proposed optimal identification (OI)
\cite{GeremiaOIShort,GeremiaOILong} procedure provides a new
approach to quantum system Hamiltonian identification through
laboratory data inversion.  The operating principle behind OI is
to improve the information content of the data by driving the
quantum system using a tailored control field (e.g., a shaped
laser pulse).  If suitably chosen, the control field forces the
data to become highly sensitive to otherwise inaccessible portions
of the Hamiltonian and therefore enables a high fidelity
inversion.  A key component of the OI concept exploits the fact
that the inversion will generally produce a family, or
distribution, of Hamiltonians that are consistent with the
laboratory data \cite{GeremiaInverseMap,GeremiaRealistic}.  The
breadth of this inversion family provides a figure of merit for
the quality of the inversion.  The limiting experimental factors
that prevent the inverted family of Hamiltonians from collapsing
down to a single (i.e., completely certain) member arise from two
sources.  First, the finite precision of the data reduces the
resolving power of the measurements and makes it possible for
multiple Hamiltonians to be consistent with the data to within its
experimental error. Second, Hamiltonians that differ in ways for
which the data is insensitive also reduces the inversion quality.
OI operates by attempting to drive the quantum system into a
dynamical state where the associated experimental errors are least
compromising, yet where the measurements provide distinguishing
power between Hamiltonians to narrow down the family consistent
with the data.

As originally formulated, the OI algorithm requires performing a
large number of global, nonlinear data inversions, each of which
can be computationally expensive \cite{GeremiaOILong}.  Resolving
the members of the inversion family consistent with the data might
require extracting hundreds of distinct Hamiltonians involving
numerous solutions of the Schr\"{o}dinger equation. In this paper,
we demonstrate that it is possible to greatly reduce the
computational challenge of performing OI by incorporating
map-facilitated inversion techniques\cite{GeremiaInverseMap,
GeremiaRealistic,GeremiaArHCl} that have been specifically
developed for efficiently finding these solution families.  A map
is a predetermined quantitative input$\to$output relationship
which can alleviate the expense of repeatedly solving the
Schr\"{o}dinger equation \cite{GeremiaForwardMap}.

This paper provides a detailed description of the algorithm
demonstrated in Ref. \cite{GeremiaOIShort} for extracting
\textit{both} internal Hamiltonian and transition dipole moment
matrix elements from simulated laser pulse shaping and population
data. Section \ref{Section::Algorithm} reviews the OI concept
introduced in Ref.\ \cite{GeremiaOILong} and then extends this
procedure to incorporate map-facilitated inversion.  Section
\ref{Section::Illustration} provides a detailed description and
in-depth analysis of the simulations that were presented in Ref.
\cite{GeremiaOIShort}.

\section{Algorithm} \label{Section::Algorithm}

The principal behind the OI algorithm is to optimize an external
control field that drives the system in a manner similar to the
learning-loop techniques utilized in many current coherent quantum
control experiments.  The distinction between OI and the latter
control experiments lies in the optimization target: OI is guided
to optimize the extracted Hamiltonian information.  The OI
learning algorithm programs a control pulse-shaper to drive the
quantum system followed by the collection of associated dynamical
observations.  In practice, the control optimization is performed
over a discrete set of variables (control ``knobs''), $\mathbf{c}
\equiv \{c_1,\ldots,c_{N_c}\}$,
\begin{equation}
    E(t) \to E(t ; c_1,c_2,\ldots,c_{N_c})
\end{equation}
where the space of accessible fields is defined by varying each
$c_i$ over a range , $c_i^{(min)}\le c_i \le c_i^{(max)}$.  For
each trial field, $E_k(t;\bc_k)$, $k=1,2,\ldots$, a set of
measurements, $\phiok$, are performed on the system. Each trial
field yields $M$ individual measurements (e.g., the populations of
$M$ different quantum states), $\phio_k=\{\phio_{k,1}, \ldots,
\phio_{k,M}\}$, with associated errors, $\{\phierr_{k,1},
\ldots,\phierr_{k,M} \}$.

Inversion is performed by adopting a discrete set of variables,
$\bh=\{h_1,\ldots,h_{N_h}\}$ used to distinguish one trial
Hamiltonian from another.  There are many possible ways to define
these Hamiltonian variables, and the best representation must be
selected to suit the quantum system being inverted, but in
general, a sufficiently flexible and accurate description of
Hamiltonian space requires a large number of variables, $N_h\gg
1$.  Inversion is accomplished by minimizing,
\begin{widetext}
\begin{equation} \label{Equation::InvCost}
        \mathcal{J}_\mathrm{inv}(\bh;\phiok)  =
    \frac{1}{M} \sum\limits_{m=1}^M \left\{
            \begin{array}{c@{\quad:\quad}l} 0 & |\phioki- \phicki | \le
            \phierrki \\
            \left\| \frac{\phioki - \phicki}{\phioki} \right\|^2
            & |\phioki - \phicki | > \phierrki \end{array} \right.
            + \hat{K}\bh
\end{equation}
\end{widetext}
where $\phicki$ is the $m^{th}$ ($m=1,\ldots,M$) observable's
computed value for the trial Hamiltonian, $\bh$, under the
influence of the external field, $E_k(t)$.  Optionally, a
regularization operator, $\hat{K}$, acting on the Hamiltonian,
$\bh$, can be used to incorporate \textit{a priori} behavior, such
as smoothness, proper asymptotic behavior, symmetry, \etc, into
the inverted Hamiltonian
\cite{Miller,HoScatter1,GeremiaInverseMap}.  While the data error
distributions are assumed to have hard bounds, $\phierrki$, in
\refeqn{InvCost}, other distributions could be used as well.

The output of the inversion optimization is a set of $N_s$
Hamiltonians, $\{\bhs_1,\ldots,\bhs_{N_s}\}$, that each ideally
reproduce the measured observable, $\phiok$, to within its
experimental error.  The upper and lower bounds of each inverted
variable defines the family,
\begin{eqnarray}
        \hsimin & = & \min\limits_s \{h_{s,i}^{*}\}
        \label{Equation::SolutionFamilyMin}\\
        \hsimax & = & \max\limits_s \{h_{s,i}^{*}\}
        \label{Equation::SolutionFamilyMax}
\end{eqnarray}
where $h_{s,i}^{*}$ is the $i^{th}$ Hamiltonian variable from the
$s^{th}$ member of $\bHs$. The uncertainty in each Hamiltonian
variable, $\hsiunc$, is quantified by the width of its
corresponding solution space,
\begin{equation} \label{Equation::SolutionError}
    \hsiunc = \hsimax - \hsimin
\end{equation}
and the width of the family for each Hamiltonian variable is used
to compute the uncertainty in the full inversion, $\Delta
\bHs[\,E_k(t)\,]$,
\begin{eqnarray} \label{Equation::ErrorMetric}
    \Delta\bHs[\,E_k(t)\,]&  = & \frac{1}{N_s}  \sum_{s=1}^{N_s}
    \mathcal{J}_\mathrm{inv}[\bhs_s;\phiok] + \\
    & & \alpha \frac{1}{N_h}
        \sum_{i=1}^{N_h}
    \left|\, \frac{2 \hsiunc}{\hsimin + \hsimax } \, \right|
    \nonumber
\end{eqnarray}
where $\bhs_s$ is the $s^{th}$ member of the inversion family
found from $E_k(t)$ and $\mathcal{J}_\textrm{inv}$ is given by
\refeqn{InvCost}. The first term in \refeqn{ErrorMetric} measures
the ability of the inversion family to reproduce the data and the
second measures the inversion uncertainty with $\alpha > 0$ being
a coefficient that balances them.

This measure of inversion uncertainty is used to guide the control
optimization where the objective is to optimize $\Delta \bHs$ over
the space of accessible fields by minimizing the control cost
function,
\begin{eqnarray} \label{Equation::ControlCost}
    \mathcal{J}_{c}[\,E(t;\bc)\,] & = & \Delta \bHs[\,E(t;\bc)\,]  + \\
    & &
    \beta \sum_{i=1}^{N_c} \left|
    \frac{c_i-c_i^{(min)}}{c_i^{(max)}-c_i^{(min)}} \right|
    \nonumber
\end{eqnarray}
where the first term reduces inversion error and the second
removes extraneous field components relative to their minimum
value, balanced by $\beta>0$.

The result of the control field optimization is a set of
laboratory data and its corresponding inversion results,
$\{E_k^*(t),\phioks,\varepsilon^*_k, \mathbf{h}^*_k\}$ that
provides the best possible knowledge of the unknown Hamiltonian
provided by the set of accessible optimal control fields.  This
inversion result provides the \textit{Optimal Identification} of
the quantum system with uncertainty,
\begin{equation} \label{Equation::OIUnc}
      \hissunc = \, \hissmax - \hissmin
\end{equation}
where $\hissmax$ and $\hissmin$ are computed using
\refeqs{SolutionFamilyMin}{SolutionFamilyMax} for the optimal
data.

The most expensive operation in OI is the many solutions of
Schr\"{o}dinger's equation required in the inversion component of
the algorithm.  The map-facilitated inversion aims to alleviate
this costly operation.  Computing quantum observables, $\phick$,
from a given Hamiltonian, $\mathbf{h}$, and control field,
$E_k(t;\mathbf{c}_k)$, defines a forward map,
\begin{equation}
    f\, :\, (\mathbf{h} ; E_k) \rightarrow \phick
\end{equation}
that is parameterized by the control field, $E_k(t)$.  In the
initial presentation of OI, this map is explicitly evaluated each
time a trial Hamiltonian is tested to determine if it is
consistent with the laboratory data. However, it has recently been
found that it is possible to pre-compute this map to high accuracy
by sampling $f(\bh;E_k)$ for a representative collection of
Hamiltonians.

Although it is generally impossible to resolve $f$ on a full grid
in $\mathbf{h}$-space due to exponential sampling complexity in
$N_h \gg 1$ dimensions, it has been found \cite{GeremiaForwardMap}
that an accurate nonlinear map can often be constructed using the
functional form,
\begin{eqnarray}
    f(\mathbf{h};E_k) & = & f_0(E_k) + \sum_{i=1}^{\hdim} \fhi
    + \sum_{i<j}^{\hdim} \fhij \nonumber  \\
          &  & + \cdots + \fhlast \label{Equation::Expansion}
\end{eqnarray}
where $f_0(E_k)$ is a constant term, the functions, $\fhi$,
dependent upon a single variable, $h_i$, the functions $\fhij$
depend upon two variables, $h_i$ and $h_j$, \etcend. Expansions of
the form in \refeqn{Expansion} belong to a family of multivariate
representations used to capture the input$\to$output relationships
of many high-dimensional physical systems\cite{AlisHDMR1,
RabitzHDMR1, ShorterHDMR1, ShorterHDMR2,ShimHDMR1, ShimHDMR2,
ShimHDMR3, GeremiaForwardMap, GeremiaControlMap}. It has been
shown that \refeqn{Expansion} converges to low order, $\mapord \ll
\hdim$ for many Hamiltonian$\to$observable maps.  A low order,
converged map expansion can be truncated after its last
significant order without sacrificing accuracy or nonlinearity
which dramatically reduces the computational labor of constructing
the map.  The specifics for how the individual expansion functions
should be evaluated can be found in previous
papers\cite{GeremiaForwardMap,GeremiaArHCl,GeremiaControlMap}.

Map-facilitated OI operates by replacing the step of explicitly
computing $\phick$ in \refeqn{InvCost} with the value $f(\bh;E_k)$
obtained by evaluating the functional map in \refeqn{Expansion}
which serves as a high-speed replacement for solving the
Schr\"{o}dinger equation. The map (or potentially multiple maps if
this is necessary to ensure sufficient accuracy
\cite{GeremiaInverseMap}) must be constructed as a preliminary
step prior to initiating the inversion optimization.  Since map
construction requires knowledge of the control field, it may not
be possible to pre-compute all of the maps prior to executing the
full OI loop.  Instead, a new map will generally need to be
constructed for each trial laser pulse.

\begin{figure*}[t]
\begin{center}
\includegraphics{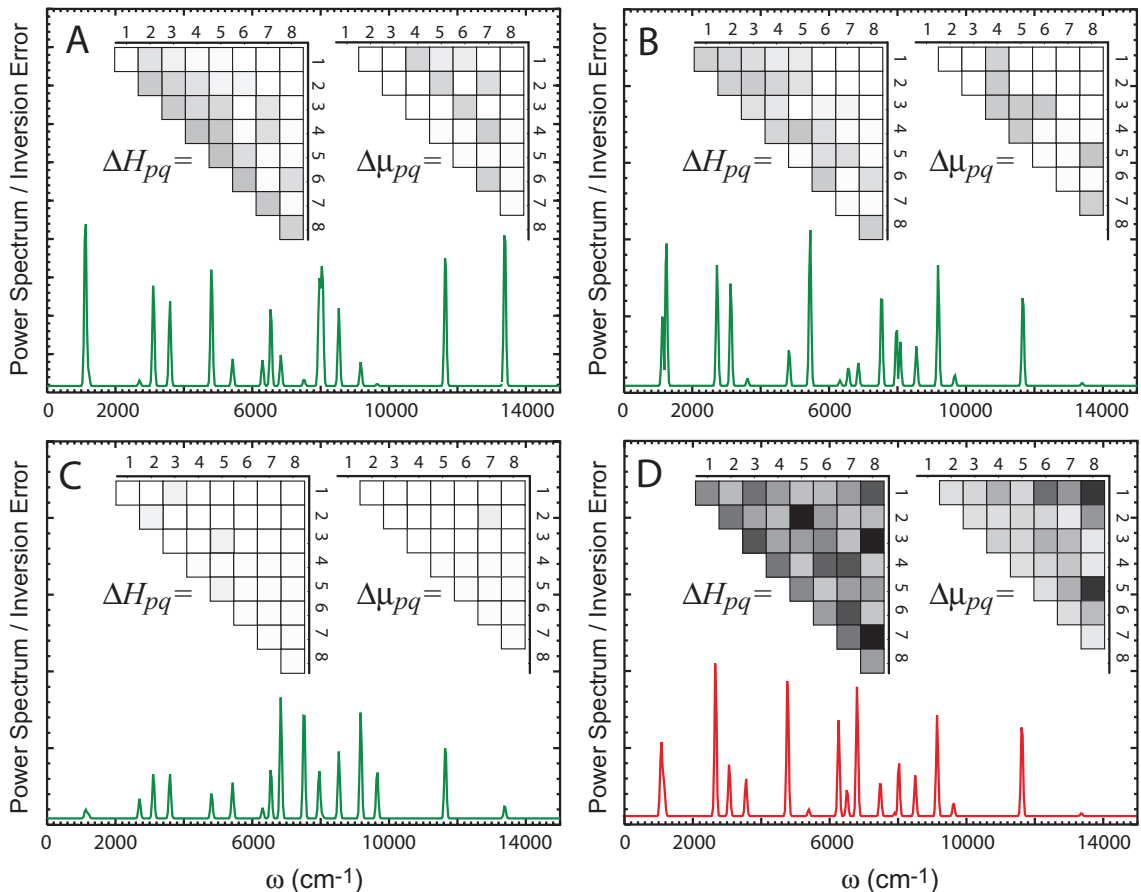}
\end{center}
\vspace{-0.8cm} \caption{(color online) Comparison of optimal
versus conventional Hamiltonian identification. In plots (A-C) the
identification error (darker shading implies larger inversion
error) respectively reflects OI's performed using 1, 2, and 4
samples of the populations during the wavepacket evolution.
Increasing the number of data points available to the OI process
improves the quality of the extracted Hamiltonian. Plot (D)
represents a conventional identification where the populations
were sampled 25 times during the time evolution. Despite using
significantly less data, OI obtained higher quality Hamiltonian
information. \label{Figure::SampleError}}
\end{figure*}

\section{Illustration} \label{Section::Illustration}

The map-facilitated OI algorithm was simulated for an 8--level
Hamiltonian\cite{8State} chosen to resemble vibrational
transitions in a molecular system where the objective was to
extract optimal information about the molecular Hamiltonian, $H$,
and dipole moment, $\mu$, for a system having the total
Hamiltonian,
\begin{equation}
    \hat{H} = H - \mu \cdot E_k(t) \,.
\end{equation}
The control fields have the form,
\begin{equation} \label{Equation::ControlPulse}
        E_k^{(j)}(t) = \exp \left( \frac{-(t-T/2)^2}{
        2 s^2}\right) \sum_l A_l^{(j)}
        \cos( \omega_l t + \theta_l^{(j)} )
\end{equation}
where the $\omega_l$ are the resonance frequencies\cite{8State} of
$H$, $A_l^{(j)}$, their corresponding amplitudes and
$\theta_l^{(j)}$, their associated phases.  Control field noise
was modeled as parametric uncertainties in the $A_l$ and
$\theta_l$,
\begin{equation}
        A_l^{(j)} = (1+\gamma_{A_l}^{(j)}) A_l, \hspace{0.5cm}
        \theta_l^{(j)} = (1+\gamma_{\theta_l}^{(j)}) \theta_l
\end{equation}
where different random values between $\pm \varepsilon^{(fld)}$
were chosen for $\gamma_{A_l}^{(j)}$ and $\gamma_{\theta_l}^{(j)}$
for each pulse.

The OI simulation involved learning the matrix elements of the
Hamiltonian, $H_{pq}= \langle p| H | q \rangle$, and dipole
moment, $\mu_{pq}=\langle p | \mu | q \rangle$ for a chosen basis,
\ket{p}, $p=1,\ldots,8$ \footnote{In practice, the basis functions
could be any complete set appropriate for the system.}. Simulated
laboratory data was generated by propagating the initial
wavefunction under the influence of the applied control field from
its initial state and computing the populations (in the chosen
basis, \ket{p}) at various times, $t_q$, during the evolution.
Each population ``measurement'' was averaged over $D=100$
replicate observations for a collection of noise-contaminated
fields, $\{E_k^{(j)}(t)\}$, $j=1,\ldots,D$, centered around the
nominal field, $E_k(t)$ to simulate experimental uncertainty in
the laser pulse-shaping process. Measurement error, $\phierr$, was
introduced into the population observations according to,
\begin{equation}
    \phioki = \Big\langle (1+\rho_{ij})\, p_i \big( t_q;
    [E_k^{(j)}(t)] \big) \Big\rangle_{j=1,\ldots,D}
\end{equation}
where the $\rho_{ij}$ were chosen randomly between
$\pm\varepsilon^{(obs)}$, the relative error in each population
observation.  A different random value, $\rho_{ij}$, was selected
for every simulated measurement.

Equation (\ref{Equation::ControlCost}) was minimized over
$\bc=\{A_l,\theta_l\}$ using a steady state GA with a population
size of 30, a mutation rate of 5$\%$, and a cross-over rate of
75$\%$. The pulse parameters where chosen to be $T=1.0$ ps and
$s=200$ fs, the amplitudes, $A_l$, were allowed to vary over [0,1]
V/\AA, and the phases were allowed to vary over [0,2$\pi$] rad.
The laboratory measurement error was assumed to be $\phierrk=2\%$,
the field error, $\varepsilon^{(fld)}=1\%$ and the observation
times, $t_q$, were uniformly spaced over the evolution period
(\ie, the time between observations was $\Delta t = T/Q$ with $Q$
= 1, 2 or 4). At each time, the full set of 8 populations was
measured.  The parameters, $\alpha$ in \refeqn{ErrorMetric} and
$\beta$ in \refeqn{ControlCost}, were ramped from \sci{1}{-4} to
\sci{1}{-2} over the GA evolution, although the optimization was
insensitive to the exact choices of $\alpha$ and $\beta$.
Typically $\sim$50 generations, or approximately 800 trial fields,
were needed for GA convergence.

Global inversion to identify the Hamiltonian family corresponding
to the data, $\phiok$, was performed by minimizing
\refeqn{InvCost} using a map-facilitated inversion algorithm. For
each inversion, the family of consistent Hamiltonians was
identified using a steady-state GA with a cross-over rate of
70$\%$ and a mutation rate of 5$\%$.   The trial family size was
$N_s=500$ and the GA population size was $N_p=100$. The
Hamiltonian-space map variables, $h_i$, were the matrix elements
of the molecular Hamiltonian, $H_{mn}$, and the the dipole,
$\mu_{mn}$.  For the eight-level system, there were 36 Hamiltonian
elements (symmetric, upper-triangle including the diagonal) and 28
transition dipole moments (symmetric, upper-triangle without the
diagonal) producing an $\hdim=64$-dimensional map.  All maps were
constructed to first order, $L=1$, and $S=6$ sample points were
used to resolve each map function for interpolation. The
Hamiltonian-space domain extended $\pm 30\%$ around its nominal
value (each matrix element was assumed known to $\pm 15\%$ prior
to the present identification).  Typical map construction required
an average of 84 seconds to perform on an SGI MIPS single
processor machine, and map-facilitated inversion required an
average of 51 seconds to converge.  A single evaluation of the
Hamiltonian$\to$population map typically required $\sim 1$ ms,
while a similar solution of the Schr\"{o}inger equation for this
system took $\sim 2$ s.  This difference is the origin of the
savings associated with map-facilitated OI.

The performance of the map-facilitated OI algorithm was assessed
with the following four tests:

\begin{enumerate}
\item[(A)] An OI was performed using populations measured at
      $Q=1$ time, $t_1=T$, producing 8 observations for the 64
      unknowns.

\item[(B)] An OI was performed using populations measured at $Q=2$
times, $t_1=T/2$ and $t_2=T$, producing 16 observation for the 64
unknowns.

\item[(C)] An OI was performed using populations measured at $Q=4$
times, $t_q=qT/4$, $q=1,\ldots,4$, producing 32 observations for
the 64 unknowns.

\item[(D)] A conventional inversion (with a randomly selected
field) was performed using populations sampled at $Q=25$ times,
$t_q=qT/25$, $q=1,\ldots,25$, producing 200 observations for the
64 unknowns.
\end{enumerate}

The power spectra of the optimal control fields for the $Q=$ 1, 2,
and 4 OI inversions are shown in \reffig{SampleError} along with a
graphical depiction of the OI error, $\hssunci$.  For a single
time sample, $Q=1$, the overall inversion error, computed as the
average, $2\langle\hssunci / (\hssmini+\hssmaxi)\rangle$, was
found to be $3.67\%$. \reffig{SampleError}(A) shows that the
majority of the error was contained in and around the diagonal
elements of $H$ (note that the Hamiltonian is not diagonal in the
chosen basis, \ket{p}). The average relative error in the dipole,
$0.9413\%$, was significantly smaller, and the majority of dipole
uncertainty appeared in the $v$$\to$$v+2$ elements.  The inversion
error for the two-point, $Q=2$, OI demonstration was reduced to
$2.910\%$ for the molecular Hamiltonian. Again, the majority of
the inversion uncertainty resides in the diagonal elements of $H$.
The transition dipole moment elements also improved with an
overall average uncertainty of $0.6270\%$.

The inversion error, $\hssunci$, for a simulated OI utilizing data
at $Q=4$ times, is essentially eliminated.  The average
uncertainty in both the molecular Hamiltonian elements and in the
transition dipoles is an order of magnitude smaller than the
simulated error in the data, $\phierrk$ = 2 \%. The most dramatic
demonstration of the OI's capabilities is seen by comparing
\reffig{SampleError}(C and D). The plot in (D) represents a
conventional inversion, performed using a randomly selected field
and $Q=25$ time samples, compared to $Q=4$ for OI. The
conventional inversion therefore had access to $M=200$ data points
while the map-facilitate OI demonstration only had $M$ = 32.
Despite what would appear to be a significant advantage in the
amount of available data, the conventional inversion displays
greater than two orders of magnitude more error. The conventional
dipole moment inversion produced an average uncertainty of
$19.5\%$ and the precision in the molecular Hamiltonian was
$28.1\%$.  The map-facilitated OI found a control field and
associated data that essentially prevented the laboratory noise
from propagating into the identified Hamiltonian information.

\section{Conclusion}

We have presented simulated experimental data that demonstrates
the utility of map-facilitate Optimal Identification for the
real-time laboratory identification of quantum Hamiltonians from
dynamical physical observable data. The central concept behind OI
is that by suitably driving the quantum system in an optimal
manner, it is possible to extract high-precision information about
all relevant aspects of its Hamiltonian despite finite laboratory
error in both the external control fields and measured data. In
this demonstration of map-facilitated OI, it was possible to
improve the computational efficiency of data inversion by more
than an order of magnitude compared to the initial presentation of
the OI concept \cite{GeremiaOIShort}.  This increased efficiency
is ultimately expected to aid, if not be essential, for the
practical implementation of OI.

\begin{acknowledgements}
This work was supported by the Department of Energy.  JMG
acknowledges support from a Princeton Plasma Science and
Technology fellowship program.
\end{acknowledgements}

\end{document}